# Dual-optical-comb spectroscopic ellipsometry

Takeo Minamikawa,[1,2,*,¶] Yi-Da Hsieh,[1,2,*] Kyuki Shibuya,[1,2] Eiji Hase,[1,2] Yoshiki Kaneoka,[1] Sho Okubo,[2,3] Hajime Inaba,[2,3] Yasuhiro Mizutani,[2,4] Hirotsugu Yamamoto,[2,5] Tetsuo Iwata,[1,2] and Takeshi Yasui[1,2]

[1]Graduate School of Science and Technology, Tokushima University, 2-1 Minami-Josanjima, Tokushima, Tokushima 770-8506, Japan

[2]Japan Science and Technology Agency (JST), ERATO Intelligent Optical Synthesizer (IOS) Project, 2-1 Minami-Josanjima, Tokushima, Tokushima 770-8506, Japan

[3]National Metrology Institute of Japan (NMIJ), National Institute of Advanced Industrial Science and Technology (AIST), 1-1-1 Umezono, Tsukuba, Ibaraki 305-8563, Japan

[4]Graduate School of Engineering, Osaka University, 2-1 Yamadaoka, Suita, Osaka 565-0871, Japan

[5]Center for Optical Research and Education, Utsunomiya University, 7-1-2 Yoto, Utsunomiya, Tochigi 321-8585, Japan

\* These authors contributed to this work equally.

¶ Correspondence and requests for materials should be addressed to

minamikawa.takeo@tokushima-u.ac.jp

**Abstract**

Spectroscopic ellipsometry is a means to investigate optical and dielectric material responses. Conventional spectroscopic ellipsometry has trade-offs between spectral accuracy, resolution, and measurement time. Polarization modulation has afforded poor performance due to its sensitivity to mechanical vibrational noise, thermal instability, and polarization wavelength dependency. We equip a spectroscopic ellipsometer with dual-optical-comb spectroscopy, viz. *dual-optical-comb spectroscopic ellipsometry* (DCSE). The DCSE directly and simultaneously obtains amplitude and phase information with ultra-high spectral precision that is beyond the conventional limit. This precision is due to the automatic time-sweeping acquisition of the interferogram using Fourier transform spectroscopy and optical combs with well-defined frequency. Ellipsometric evaluation without polarization modulation also enhances the stability and robustness of the system. In this study, we evaluate the DCSE of birefringent materials and thin films, which showed improved spectral accuracy and a resolution of up to $1.2 \times 10^{-5}$ nm across a 5-10 THz spectral bandwidth without any mechanical movement.

**Introduction**

Ellipsometry is widely used for the investigation and evaluation of materials in terms of optical and dielectric response in both academic and industrial research. An ellipsometer measures the polarization state of light incident upon a material to determine the material properties, such as complex dielectric function, carrier structure, crystalline nature, and thickness of a thin film[1-3]. Spectroscopic evaluation of ellipsometry, namely spectroscopic ellipsometry, provides further insight into materials, not only for spectral characteristics but also for estimation of properties that cannot be determine by conventional ellipsometry alone[1]. Spectroscopic ellipsometry measures the complex reflectance ratio $\rho$,

$$\rho = \tan(\Psi) \exp(i\Delta), \qquad (1)$$

where $\Psi$ and $\Delta$ respectively denote the amplitude ratio and phase difference between *p*- and *s*-polarization components of the polarization state of incident light, the so-called ellipsometric parameters. These ellipsometric parameters are obtained as a function of wavelength. In conventional spectroscopic ellipsometry, the ellipsometric parameters are obtained from the optical intensity measurements resulting from the modulation of the polarization state of light interacting with the target material; this limits the mechanical stability, thermal stability, and spectral bandwidth of simultaneous spectroscopic observation. Furthermore, the spectral resolution and accuracy obtained in spectroscopic ellipsometry are limited due to the spectral resolution limit of conventional dispersive spectrometers or the mechanical time-sweep nature of conventional Fourier transform spectrometers. The practical spectral resolution and accuracy of conventional ellipsometry are, thus, limited to a few tens of GHz corresponding to $10^{-1}$-$10^{-2}$ nm (1-10 GHz).

In order to overcome these limitations, we propose spectroscopic ellipsometry employing dual-optical-comb spectroscopy, namely *dual-optical-comb spectroscopic ellipsometry* (DCSE). Dual-optical-comb spectroscopy is a promising technique for ultra-precise, accurate, and broadband spectroscopy[4-19]. Dual-optical-comb spectroscopy employs two optical-comb lasers having slightly different repetition rates, allowing the acquisition of a complete interferogram with an ultra-wide time span without mechanical scanning for deducing the highly resolved optical spectrum[18]. Since the optical-comb laser is quite stable, owing to the stabilization of the repetition rate and carrier envelope offset, the wavelength of each comb can be accurately and precisely determined to within 100 kHz to 10 MHz ($10^{-6}$-$10^{-4}$ nm)[18, 20]. Furthermore, dual-optical-comb spectroscopy is based on Fourier transform spectroscopy, which provides amplitude and phase spectra by directly decoding from the interferograms of two optical-comb lasers, enabling polarization analysis in terms of amplitude and phase spectra along two orthogonal axes. These advantages of the dual-optical-comb spectroscopy enable spectroscopic ellipsometry with ultra-high spectral resolution and accuracy for deducing spectra and polarization states without any mechanical movements.

In this study, we provide a proof-of-principle demonstration of DCSE for novel material evaluation. We developed a DCSE system by using two optical-comb laser sources employing erbium-doped-fibre-based mode-locked lasers. We constructed the system with reflection and transmission configurations to demonstrate the polarization analytical capability of DCSE. DCSE evaluation of birefringent materials and thin films were successfully realized with a spectral resolution of 1.5 MHz over 5-10 THz without any mechanical movement.

**Principle operation**

In DCSE, the ellipsometric parameters are evaluated in terms of Fourier transformation of interferograms observed in *p*- and *s*-polarization components in the reflection configuration or x- and y-polarization components in the transmission configuration (Fig. 1a, also see Methods). Two highly stabilized, amplified comb lasers with repetition rates of $f_{rep,S}$ and $f_{rep,L}$ were synchronized with slightly different repetition rates ($\Delta f_{rep}=f_{rep,S}-f_{rep,L}$). The temporal interferogram of the two comb lasers had a periodic time delay interval of $\Delta T=1/f_{rep,L}-1/f_{rep,S}$. The amplitude and phase of each polarization component were decoded with the interferograms in terms of Fourier transform spectroscopy. When the light of a comb laser (signal comb laser) interacts with a sample, the resultant polarization modulation of the light is directly encoded with the amplitude and phase into the interferograms.

Ellipsometric parameters $\Psi$ and $\Delta$ were, thus, evaluated by decoding the interferograms obtained from the photodetectors as follows:

$$\Delta = \arg(I_{PD1}) - \arg(I_{PD2}) + \pi, \qquad (2)$$

$$\Psi = \tan^{-1} \frac{|I_{PD1}|}{|I_{PD2}|}. \tag{3}$$

where $I_{PD1}$ and $I_{PD2}$ indicate the individually observed signal of each polarization with photodetectors PD1 and PD2, of which amplitude ($|I|$) and phase (arg ($I$)) were decoded by Fourier transformation. Since the angle of the polarizers at the signal comb and local comb arms were respectively set at π/4 and -π/4 rad in our setup, the phase difference of each arm initially had a phase shift of π rad as shown in Eq. 2. The evaluation of $\Psi$ and $\Delta$ does not require mechanical or wavelength-dependent modulation of polarization states, indicating that scan-less spectroscopic ellipsometry with wide spectral bandwidth can be realized.

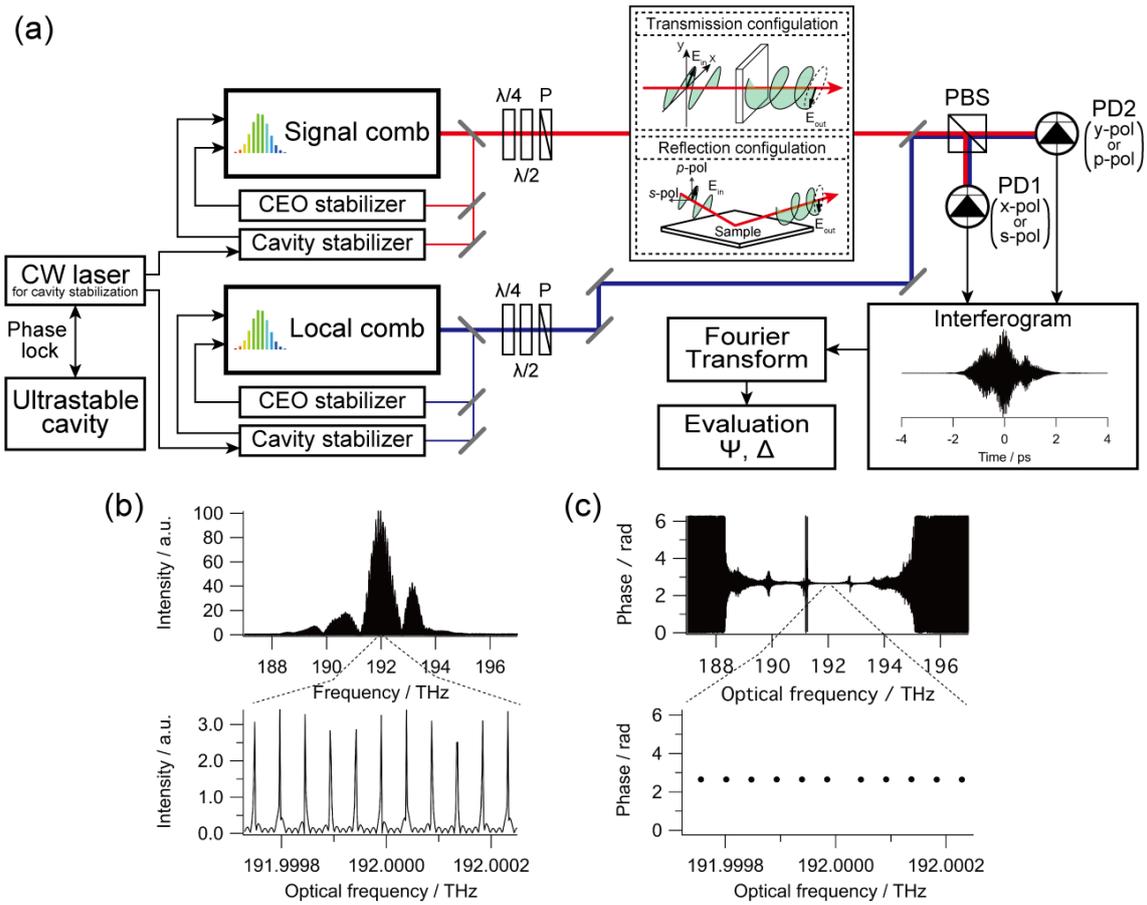

Figure 1. Experimental setup and fundamental specification of the developed DCSE system. (a) Configuration of the laser stabilization system and optical setup of the DCSE. Components include a continuous wave (CW) laser locked to an ultra-stable cavity; quarter-waveplate (λ/4); half-waveplate (λ/2); polarizer (P); polarization beam splitter (PBS); photo-detector (PD); *p*-polarization (*p*-pol); *s*-polarization (*s*-pol); x-polarization (x-pol); and y-polarization (y-pol). Fundamental specification of (b) an amplitude spectrum of *p*-polarized light and (c) a phase difference spectrum of *p*- and *s*-polarized light by applying an interferogram concatenation method with 20

duplications of a single interferogram. A spectral resolution of 1.5 MHz ($1.2 \times 10^{-5}$ nm) and a spectral spacing of 48 MHz were achieved.

**Basic spectral performance of DCSE**

Basic spectral performance of the DCSE system was evaluated with a gold mirror in the reflection configuration, as illustrated in Figs. 1b and c. By applying an interferogram concatenation method[16, 21] with 20 duplications of a single interferogram with a one complete period of the repetition rate, we achieved an ultra-high spectral resolution of 1.5 MHz ($1.2 \times 10^{-5}$ nm) and spectral spacing of 48 MHz, as shown in the amplitude spectrum in Fig. 1b. Since $\Psi$ was examined by using *p*- and *s*-polarization of each comb, the spectral resolution of $\Psi$ coincided with that of the amplitude spectra of *p*- and *s*-polarization components. The phase spectrum was constructed by extracting the phase data at the optical frequency of each comb peak of the amplitude spectrum, the spectral resolution also coincided with that of the amplitude spectrum in Fig. 1c. Furthermore, the spectral interleaving method[16, 18] or multi-interferogram observation[22], which is generally used in dual-optical-comb spectroscopy, can be used in DCSE to gain higher spectral resolution.

**Soleil-Babinet compensator**

As a first experimental demonstration, we performed an ellipsometric evaluation of a Soleil-Babinet compensator in the transmission configuration. The ellipsometric parameters of the Soleil-Babinet compensator observed with the DCSE system with a variety of relative wedge distances are shown in Fig. 2 and Supplementary Movies 1 and 2. The theoretical relationship of ellipsometric parameters and the relative wedge distance, which were calibrated at the optical frequency of 193.4 THz (1550 nm), is also shown. Since the fast axis of the Soleil-Babinet compensator was aligned along the axis of x-polarization in the sample coordinate, the Soleil-Babinet compensator worked as a relative phase retarder of the x- and y-polarization components of the signal comb laser without the amplitude modulation relative to each polarization. The ellipsometric parameters, $\Psi$ and $\Delta$, determined with the DCSE system reflected the following optical conditions. The flat structure of $\Psi$ indicated that the relative amplitude of x- and y-polarization components was insensitive to the relative wedge distance of the Soleil-Babinet compensator and the optical frequency. In addition, the linear relationship between $\Delta$ and the relative wedge distance of the Soleil-Babinet compensator was that of a phase retarder relative to the x- and y-polarization components. The DCSE estimation errors of $\Psi$ and $\Delta$ are shown in Supplementary Fig. 1. We obtained root mean square errors of 19.3 mrad and 58.0 mrad for $\Psi$ and $\Delta$, respectively, at the calibrated optical frequency of 193.4 THz (1550 nm). We obtained root mean square errors of 24.5 mrad and 109 mrad for $\Psi$ and $\Delta$, respectively, for the optical frequency range 191-195 THz. These results clearly support the assertion that DCSE has the potential for polarization analysis along two orthogonal axes for ellipsometric analysis of birefringent materials.

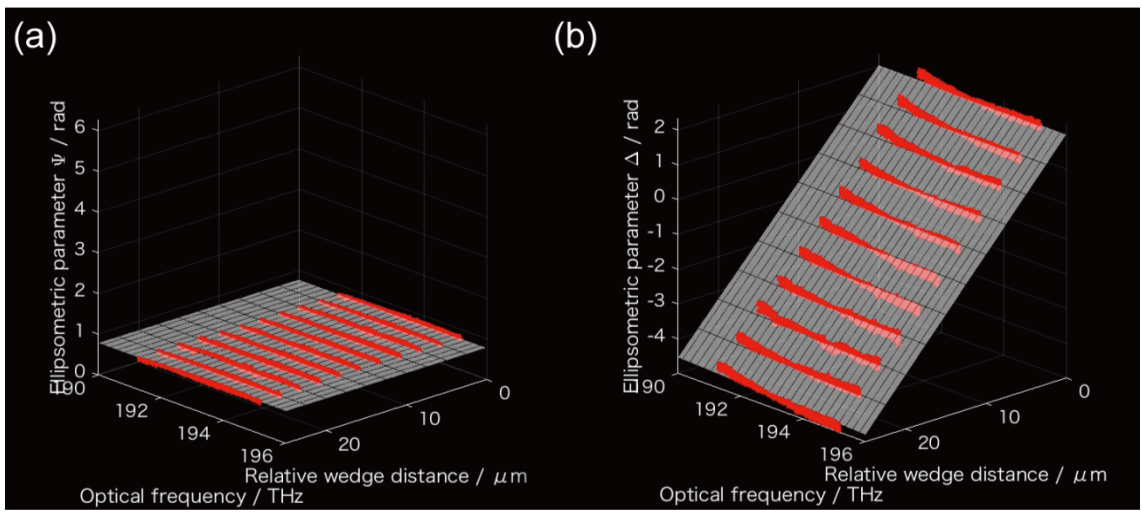

Figure 2. Ellipsometric evaluation of a Soleil-Babinet compensator with the DCSE system. Ellipsometric parameters, (a) $\Psi$ and (b) $\Delta$, obtained through DCSE (red dots) and theoretical estimation (mesh surface).

**High-order waveplates**

To evaluate the detection capability of an optical-frequency-dependent birefringent material, we performed ellipsometric evaluation of a high-order waveplate in the transmission configuration. Figure 3 and Supplementary Movies 3 and 4 show DCSE evaluation of a high-order waveplate designed as a 30th order quarter waveplate at 633 nm which also works as high order waveplate in the 1550 nm region. Among the 360º rotation of the high-order waveplate with respect to the linear polarization of the incident light, we obtained four peaks on $\Psi$ and two peaks on $\Delta$. This is in good agreement with the $\Psi$ and $\Delta$ behaviour of the incident-angle dependency of the optically uniaxial phase retardation of a general waveplate. Regarding optical frequency, we obtained the rotation-angle-insensitive region at 192.5 THz, which indicated that the phase retardation between x- and y-polarization was equal to an integer multiple of $2\pi$.

This result supports DCSE's application in ellipsometrically evaluating optical frequency-dependent birefringent materials.

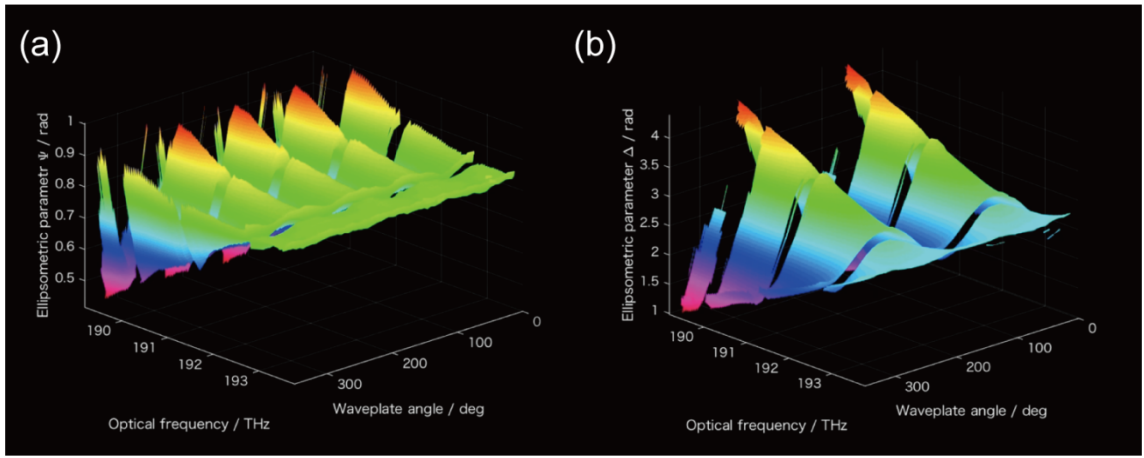

Figure 3. Ellipsometric evaluation of a high-order waveplate with the DCSE system. Ellipsometric parameters of (a) $\Psi$ and (b) $\Delta$ obtained with the DCSE system at the variety of the rotation angle of the high-order waveplate.

**$SiO_2$ thin film standards**

Finally, we applied the proposed DCSE to the ellipsometric evaluation of thin films in the reflection configuration. We performed a thickness prediction of thin film standards that had a $SiO_2$ thin film with 0 to 900 nm in thickness deposited on a Si base plate. Figures 4a and b and Supplementary Movies 5 and 6 show that $\Psi$ and $\Delta$ depend on film thickness at an incident angle of 58°. The experimental results of $\Psi$ and $\Delta$ values were in good agreement with the theoretical estimation calculated assuming a three-layer model with air/SiO$_2$/Si.

We predicted the film thickness by minimizing the root mean square error of the experimental and theoretical values of $\Delta$ over a 10 THz spectral bandwidth. The local minimum of the root mean square error was observed approximately every 650 nm due to the 10 THz spectral bandwidth used for the film thickness estimation (Supplementary Fig. 2). We estimated the film thickness by evaluating the local minimum within 500 nm of the theoretical estimates. The film thickness estimation was realized as shown in Fig. 4c and the root mean square error of the film thickness estimation was 37.9 nm over 900 nm. These results confirmed DCSE's capability to evaluate thin films.

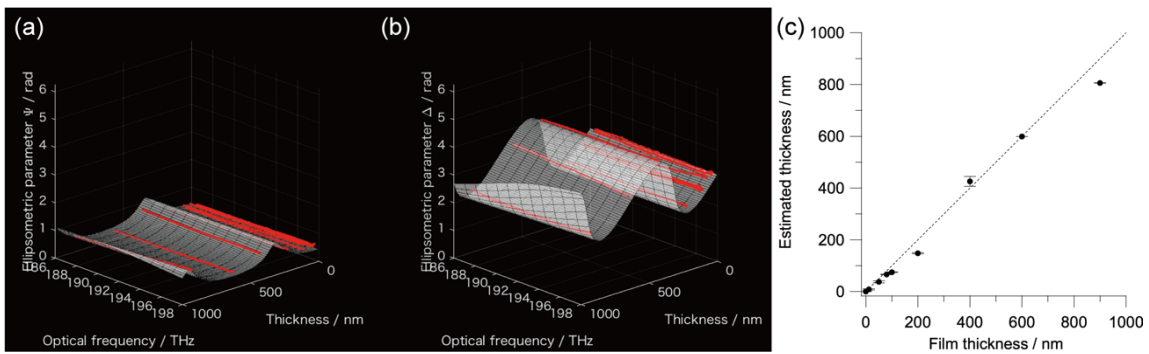

Figure 4. Ellipsometric evaluation of thin film standards with the DCSE system. Ellipsometric parameters, (a) $\Psi$ and (b) $\Delta$, obtained with the DCSE system (red dots) and by theoretical estimation (mesh surface). (c) Thickness prediction by DCSE. Error bars indicate standard deviation.

**Discussion**

This study provided a proof-of-principle demonstration of ultra-high spectral resolution DCSE without any mechanical movements in the spectrometer or polarization modulation. Our approach can directly and simultaneously observe the modulated polarization states that resulted from the interaction of polarized light with various materials, in terms of amplitude and phase spectra of two orthogonal polarizations. The dual-optical-comb polarization measurements need only the stabilization of comb sources, enabling fast and robust ellipsometric evaluation with ultra-high spectral resolution. Furthermore, a 5-10 THz (40-80 nm) spectral bandwidth was achieved in this study. The results presented confirm the feasibility of DCSE for material evaluation.

The high spectral resolution is generally required in the ellipsometric evaluation of highly wavelength-dependent materials as shown in Fig. 3. Further application is expected in the evaluation of an absorbable material with vibrational transitions in, for instance, the mid- or far-infrared region[23-25]. In these ellipsometry applications, the high spectral-resolution of the DCSE can be effectively utilized for precise and accurate evaluation of material properties. Furthermore, ellipsometric evaluation with a broadband spectrum can be employed to evaluate a material with multiple unknown parameters, e.g., a multi-layered structure or rough surface[1]. A spectral bandwidth of 5-10 THz (40-80 nm) was realized in this study, however, it can be enhanced to 190 THz (900 nm) with the same stabilization system as our setup[17]. Dual-optical-comb spectroscopy's advantageous automatic time-sweep with well-frequency-defined optical combs, allowed us to achieve broadband and high spectral resolution ellipsometric evaluation without any mechanical movements.

Measurement time for ellipsometric evaluation in DCSE depends on the frequency difference of the repetition rate of the two comb lasers ($\Delta f_{rep}$) and the number of signal averaging of interferograms. Since interferograms in DCSE are observed every $1/\Delta f_{rep}$, the measurement time with signal averaging coincides with the multiplication of $1/\Delta f_{rep}$ and the number of signal averaging. In our demonstration, we employed a 21 Hz frequency difference for the repetition rates of the two comb lasers and 1000 times signal averaging to achieve a sufficient signal-to-noise ratio. The measurement time was about 48 s for ellipsometric evaluation including polarization analysis using the amplitude and phase of two orthogonal axes and spectral decoding by means of Fourier transformation with the high spectral resolution of 1.5 MHz. Since the frequency difference of the repetition rate of the two comb lasers can be optimized in terms of spectral bandwidth for DCSE in the same way as dual-optical-comb spectroscopy[18], the optimized frequency difference of the repetition rate is estimated to be 230 Hz and 115 Hz for 5 THz and 10 THz spectral bandwidths, respectively. This result indicates that DCSE can be performed 5 to 10 times faster than our demonstration. Furthermore, the required number of signal averaging was determined by sufficient signal-to-noise ratio of interferograms, which depend on experimental conditions, such as optical throughput of system, optical-comb laser stability, detectors and spectral characteristics, such as spectral resolution and bandwidth. This indicates that faster DCSE evaluation can be achieved if we realize single-shot measurement of interferograms. Although the minimum measurement time of DCSE depends on experimental conditions, fast ellipsometric evaluation of DCSE with high spectral resolution and accuracy, and wide spectral bandwidth can be realized owing to the automatic time-sweep nature of DCSE for both the polarization analysis and spectral decoding.

In conclusion, we proposed an ellipsometric material evaluation method employing dual-optical-comb spectroscopy, namely DCSE, and provided a proof-of-principle demonstration of DCSE by using birefringent materials and thin film standards. We expect that our method will be a powerful tool for material science beyond the conventional limit of ellipsometry.

**Methods**

**Experimental setup**

The optical setup of the DCSE system is shown in Fig. 1. Two highly stable erbium-based mode-locked fibre lasers were employed as the signal and local comb oscillators, which have been described previously[17]. Each fibre laser consisted of an electro-optic modulator, a piezo-electric transducer, and a Peltier thermo-controller for the broadband and long-term stabilization of the laser cavity. A delay line was also installed for tuning the repetition rate of the pulse train in the laser cavity. The repetition rates of the two combs were set to about 48 MHz, the frequency difference of 21 Hz was precisely stabilized by phase-locking between a well-stabilized 1.54 µm continuous wave laser and the nearest comb mode. The carrier-envelope offset frequency of each comb, which were detected by using a f-2f configuration[26], was phase-locked to a reference frequency by controlling the injection current of the pump lasers for the comb oscillator. Our stabilization system of the repetition rate and carrier-envelope offset frequency achieved a narrow sub-Hz relative linewidth between combs.

      The laser outputs of the individual comb lasers for DCSE measurements were individually amplified by erbium-doped fibre amplifiers, then spectrally broadened with highly nonlinear fibres. The polarization states of the two optical-comb lasers (signal and local combs) were set at linear polarization at 45º and -45º, respectively. The signal comb was incident on the sample in reflection or transmission configuration and was

spatially overlapped with the local comb laser using a beam splitter. The interferograms of the *p*- and *s*- (or x- and y-) polarization components of the comb lasers were individually detected using a polarization beam splitter and InGaAs photo-detectors. The interferograms were averaged 1000 times in the time domain with a coherent averaging condition[27]. Furthermore, we performed the real-time phase compensation of the carrier phase drift of interferograms and the frequency drift of the continuous wave laser as described previously[17]. The interferograms were processed with a computer by fast Fourier transformation to reconstruct the amplitude and phase spectra.

**Evaluation method of ellipsometric parameters, $\Psi$ and $\Delta$, with DCSE**

The polarization state of light propagation along our system can be illustrated in terms of Jones calculus as,

$$\boldsymbol{E_{obs}} = \boldsymbol{S}\boldsymbol{P_S}(\boldsymbol{\theta_S})\boldsymbol{E_S} + \boldsymbol{P_L}(\boldsymbol{\theta_L})\boldsymbol{E_L}, \qquad (2)$$

where *E*, *S*, and *P* represent electric fields, sample, and polarizers oriented at angle *θ*. Subscripts *obs*, *S*, and *L* refer to the observed position, signal-comb arm, and local-comb arm.

Since the rotation angle of the polarizers at signal-comb and local-comb arms were set at π/4 and -π/4, respectively, and the amplitudes of the electric fields after the polarizer were $\bar{E}_L$ and $\bar{E}_S$, equation (2) can be expressed as,

$$E_{obs} = \frac{1}{\sqrt{2}} \begin{bmatrix} \bar{E}_L + \bar{E}_S \exp(-i\Delta)\sin\Psi \\ \bar{E}_L \exp(-i\pi) + \bar{E}_S \cos\Psi \end{bmatrix}, \quad (3)$$

where,

$$S = \begin{bmatrix} \exp(-i\Delta)\sin\Psi & 0 \\ 0 & \cos\Psi \end{bmatrix}. \quad (4)$$

We observed the interferograms of each polarization component by using a polarization beam splitter and extracting the cross-correlation term of the signal and local combs in equation (3). Since the Fourier transformation of the interferograms decodes the amplitude and phase of the cross-correlation terms, $\Psi$ and $\Delta$ are directly calculated with the amplitude and phase spectra as shown in equations (2) and (3).

**Samples**

The fast axis of a Soleil-Babinet compensator (SBC-IR, Thorlab, Inc.) was aligned at 45º along x-axis of the sample coordinate. To prepare the theoretical estimation of the retardation of the Soleil-Babinet compensator, the relative wedge distance of 0 to 25 mm of the Soleil-Babinet compensator was calibrated at 193.4 THz (1550 nm) by using the phase obtained with DCSE. DCSE evaluation of the Soleil-Babinet compensator was performed in 2.5 mm steps over the relative wedge distance of 25 mm. To achieve the theoretical estimation of the retardation of the Soleil-Babinet compensator, the ordinary and extraordinary refractive indices of the Soleil-Babinet compensator were estimated using a Sellmeier model whose coefficients were provided by the manufacturer (Thorlab, Inc.).

A custom-made high-order waveplate was designed as a quarter waveplate at 633 nm, which also worked as a high-order waveplate at 1550 nm, as demonstrated in this study. DCSE evaluation of the high-order waveplate was performed in 10º increments over 360º.

The thin film standards (STA-6000-C and STA-9000-C, Five Lab Co., Ltd.) were measured in the reflection configuration. The incident light angle was aligned to 58º, near the Brewster's angle of $SiO_2$ (55.2º), which was confirmed by the comparison of the theoretical estimates and observed $\Psi$ and $\Delta$ from a thickness of 0 to 900 nm of the thin film standards. Film thicknesses of 0, 12, 50, 80, 100, 200, 400, 600, and 900 nm were used for the DCSE evaluation. The theoretical estimates were calculated with a three-layer model of air/$SiO_2$/Si, where the $SiO_2$ thickness was set from 0 to 1000 nm thickness and the thicknesses of air and Si were set to infinity. The complex refractive indices of air, $SiO_2$, and Si were assumed to be 1.00, 1.46, and $3.87-i1.46 \times 10^{-2}$, respectively.

**Acknowledgement**

This work was supported by Exploratory Research for Advanced Technology (ERATO) MINOSHIMA Intelligent Optical Synthesizer Project, Japan Science and Technology Agency (JST), Japan.

**Author contributions**

T.Y. and T.I. conceived the project. T.M., Y.-D.H., T.I. and T.Y. designed the experiments. T.M., Y.-D.H., K.S., E.H., Y.K. and S.O. performed the experiments. S.O.

and H.I. constructed the dual-comb system. T.M. and Y.-D.H. analysed the data. T.M. wrote the manuscript. Y.M., H.Y., T.I. and T.Y.    contributed to manuscript preparation. All authors discussed the results and commented on the manuscript.

**Competing financial interests:**

The authors declare no competing financial interests.

comb teeth over 43 THz. *Opt. Lett.* 2012, **37**(4)**:** 638-640.

12. Roy J, Deschenes JD, Potvin S, Genest J. Continuous real-time correction and averaging for frequency comb interferometry. *Opt. Express* 2012, **20**(20)**:** 21932-21939.

13. Ideguchi T*, et al.* Coherent Raman spectro-imaging with laser frequency combs. *Nature* 2013, **502**(7471)**:** 355-358.

14. Zhang ZW, Gardiner T, Reid DT. Mid-infrared dual-comb spectroscopy with an optical parametric oscillator. *Opt. Lett.* 2013, **38**(16)**:** 3148-3150.

15. Ideguchi T*, et al.* Adaptive real-time dual-comb spectroscopy. *Nat. Commun.* 2014, **5:** 3375.

16. Yasui T*, et al.* Super-resolution discrete Fourier transform spectroscopy beyond time-window size limitation using precisely periodic pulsed radiation. *Optica* 2015, **2**(5)**:** 460-467.

17. Okubo S*, et al.* Ultra-broadband dual-comb spectroscopy across 1.0-1.9 mu m. *Appl. Phys. Express* 2015, **8**(8).

18. Coddington I, Newbury N, Swann W. Dual-comb spectroscopy. *Optica* 2016,
28